\documentclass[12pt,a4paper]{article}

\usepackage{amsmath}
\usepackage{amssymb}
\usepackage{graphicx}
\usepackage{overpic}
\usepackage{cite}
\usepackage{tabls}
\usepackage{hhline}
\usepackage{dcolumn}
\usepackage{enumitem}

\let\oldappendix=\appendix
\let\oldsection=\section
\renewcommand{\appendix}{\oldappendix%
\def\theequation{\Alph{section}.\arabic{equation}}%
\renewcommand{\section}{\setcounter{equation}{0}\oldsection}}

\newcommand{\beq}{\begin{equation}}
\newcommand{\eeq}{\end{equation}}
\newcommand{\beqa}{\begin{eqnarray}}
\newcommand{\eeqa}{\end{eqnarray}}

\newcommand{\tr}{\mbox{tr}}

\newcommand{\newop}[2]{\def#1{\mathop{\mathrm{#2}}\nolimits}}
\newop{\artanh}{artanh}
\newop{\det}{det}
\newop{\tr}{tr}
\newop{\diag}{diag}
\newop{\Re}{Re}
\newop{\Im}{Im}

\newcommand{\vs}{\vspace{-0.25cm}}

\begin{document}

\hfill {\tiny HISKP-TH-06/24, FZJ-IKP(TH)-2006-23}

\hfill 

\bigskip\bigskip

\begin{center}

{\Large\bf On the extraction of the quark mass ratio {\boldmath$(m_d-m_u)/m_s$}\\[0.3ex]
  from {\boldmath$\Gamma(\eta' \to \pi^0 \pi^+ \pi^-)/\Gamma(\eta' \to \eta \pi^+ \pi^-)$}}

\end{center}

\vspace{.4in}

\begin{center}
{\large B.~Borasoy\footnote{email: borasoy@itkp.uni-bonn.de}$^{a}$,
        Ulf-G.~Mei{\ss}ner\footnote{email: meissner@itkp.uni-bonn.de}$^{a, b}$,
        R.~Ni{\ss}ler\footnote{email: rnissler@itkp.uni-bonn.de}$^{a}$}

\bigskip

\bigskip

$^{a}$Helmholtz-Institut f\"ur Strahlen- und Kernphysik (Theorie), \\
Universit\"at Bonn,
D-53115 Bonn, Germany \\[0.4cm]

$^b$Institut f\"ur Kernphysik (Theorie), Forschungszentrum J\"ulich, \\
D-52425 J\"ulich, Germany \\[2ex]

\vspace{.2in}

\end{center}

\vspace{.7in}

\thispagestyle{empty} 

\begin{abstract}
\noindent The claim that the light quark mass ratio $(m_d - m_u)/m_s$ can be
extracted from the decay width
ratio $\Gamma(\eta' \to \pi^0 \pi^+ \pi^-)/\Gamma(\eta' \to \eta \pi^+ \pi^-)$
is critically investigated within a U(3) chiral unitary framework. 
The influence of
the recent VES data on the $\eta' \to \eta \pi^+ \pi^-$ decay is 
also discussed.
\end{abstract}\bigskip

%\begin{center}
%\begin{tabular}{ll}
%\textbf{PACS:}& 11.80.-m, 12.39.Fe, 13.75.Jz, 36.10.Gv \\[6pt]
%\textbf{Keywords:}& Chiral Lagrangians, coupled channels, unitarity.
%\end{tabular}
%\end{center}

% 11.30.Rd Chiral symmetries
% 12.39.Fe Chiral Lagrangians
% 12.40.Vv Vector-meson dominance
% 13.20.Cz Decays of pi mesons
% 13.40.Hq Electromagnetic decays
% 36.10.Gv Mesonic atoms and molecules, hyperonic atoms and molecules
% 13.75.Gx Pion-baryon interactions
% 13.75.Jz Kaon-baryon interactions
% 11.80.-m Relativistic scattering theory
% 11.80.Gw Multichannel scattering

\vfill

%%%%%%%%%%%%%%%%%%%%%%%%%%%%%%%%%%%%%%%%%%%%%%%%%%%%%%%%%%%%%%%%%%%%%%%%%%%%%%%%
%%%%%%%%%%%%%%%%%%%%%%%%%%%%%%%%%%%%%%%%%%%%%%%%%%%%%%%%%%%%%%%%%%%%%%%%%%%%%%%%
\section{Introduction} \label{sec:intro}
%%%%%%%%%%%%%%%%%%%%%%%%%%%%%%%%%%%%%%%%%%%%%%%%%%%%%%%%%%%%%%%%%%%%%%%%%%%%%%%%

The light quark masses $m_u, m_d, m_s$ are fundamental parameters of Quantum Chromo
Dynamics and ought to be constrained as accurately as possible. The determination
of the light quark mass ratios has been the goal of a variety of investigations
in low-energy hadron physics, see e.g. 
\cite{Gasser:1982ap, Leutwyler:1989pn, Donoghue:1992ac, Leutwyler:1996qg, Amoros:2001cp}. 
Of particular interest is the quark mass difference
$m_d - m_u$ which induces isospin breaking in QCD. Moreover, the possibility $m_u =0$ would provide
an explanation for the strong $CP$ problem.

An accurate way of extracting $m_d - m_u$ is given by the isospin-violating
decays $\eta,\eta' \to \pi^0 \pi^+ \pi^-$ and $\eta, \eta' \to  3 \pi^0$.
While for most processes isospin-violation of the strong interactions is
masked by electromagnetic effects, these corrections are expected to be small for the
three pion decays of the $\eta$ and $\eta'$ (Sutherland's theorem) \cite{Sutherland:1966mi} 
which has been confirmed in an effective Lagrangian framework \cite{BKW}.
Neglecting electromagnetic corrections the decay amplitudes are directly proportional to 
$m_d-m_u$.

For this reason, it has been claimed in  \cite{GTW} that the branching ratio
$r = \Gamma(\eta' \to \pi^0 \pi^+ \pi^-)/\Gamma(\eta' \to \eta \pi^+ \pi^-)$
can be utilized in a very simple manner
to extract the light quark mass difference $m_d - m_u$.
To this aim, it is assumed that
\begin{enumerate}[label=\emph{\alph*}),ref=\emph{\alph*}]
\item \label{assuma} the amplitude $A(\eta' \to \pi^0 \pi^+ \pi^-)$ is determined by
  the corresponding amplitude $A(\eta' \to \eta \pi^+ \pi^-)$ via
  \beq  \label{eq:amprel}
  A(\eta' \to \pi^0 \pi^+ \pi^-) = \epsilon \ A(\eta' \to \eta \pi^+ \pi^-)
  \eeq
  with $\epsilon = (\sqrt{3}/4) \, (m_d - m_u)/(m_s - \hat{m})$ 
  the $\pi^0$-$\eta$ mixing angle and $\hat{m} = (m_d + m_u)/2$.
  (Note that in \cite{GTW} the difference $m_s - \hat{m}$ has been approximated by $m_s$ in the 
  denominator of $\epsilon$.)
  Eq.~(\ref{eq:amprel}) implies that the decay $\eta' \to \pi^0 \pi^+ \pi^-$ proceeds entirely 
  via $\eta' \to \eta \pi^+ \pi^-$ followed by  $\pi^0$-$\eta$ mixing.
\item \label{assumb} both amplitudes are \emph{``essentially constant''} over phase space
  (see the remark in front of Eq.~(19) of Ref.~\cite{GTW}).
\end{enumerate}
Based on these two assumptions one arrives at the relation
\beq \label{eq:gtw} 
r = \frac{\Gamma(\eta' \to \pi^0 \pi^+ \pi^-)}{\Gamma(\eta' \to \eta \pi^+ \pi^-)}
\simeq (16.8)\frac{3}{16}\left(\frac{m_d - m_u}{m_s}\right)^2 ,
\eeq
where the factor $16.8$ represents the phase space ratio.
Comparison with experimental data---for which, so far, only an upper limit 
exists---would then lead to a prediction for the quark mass ratio 
$(m_d - m_u)/(m_s - \hat{m}) \simeq (m_d-m_u)/m_s$.
The purpose of the present work is to critically examine these two assumptions
which lead to the simple relation in Eq.~(\ref{eq:gtw}).
Such an investigation is very timely in view of the recent and ongoing experimental
activities on $\eta$ and $\eta'$ decays at the WASA facility at COSY \cite{WASA},
MAMI-C \cite{Nef}, KLOE at DA$\Phi$NE \cite{KLOE} and by 
the VES Collaboration \cite{VES1, VES2}.

An appropriate theoretical framework to investigate low-energy hadronic physics is
provided by chiral perturbation theory (ChPT) \cite{GL}, the effective field theory 
of QCD. In ChPT Green functions are expanded perturbatively in powers of Goldstone boson
masses and small momenta. 
However, final-state interactions in $\eta \to 3 \pi$ 
have been shown to be substantial both in a complete one-loop calculation
in SU(3) ChPT \cite{GL2} and using a dispersive framework \cite{Kambor, Anisovich:1996tx}.
It is hence important to include final-state interactions in a non-perturbative fashion.

In $\eta'$ decays final-state interactions are expected to be even more important
due to larger phase space and the presence of nearby resonances.
In this investigation, we apply the framework of U(3) chiral effective field theory
in combination with a relativistic coupled-channels approach developed in \cite{BB1, BN1}
in order to calculate the ratio $r$.
Final-state interactions are included by deriving $s$- and $p$-wave interaction
kernels for meson-meson scattering from the chiral effective Lagrangian and
iterating them in a Bethe-Salpeter equation. 
The infinite iteration of the chiral effective potentials generates resonances dynamically.
Very good overall agreement with currently available data on $\eta$, $\eta'$ decay widths and 
spectral shapes has been achieved in \cite{BB1, BN1}.

In the next section, we will investigate in our approach if both the assumptions ``\ref{assuma}''
and ``\ref{assumb}''
are justified. The inclusion of the recent VES data \cite{VES2} which provide
higher statistics on the spectral shape of $\eta' \to \eta  \pi^+ \pi^-$ than previous experiments
is studied in Sec.~\ref{sec:VES}, while Sec.~\ref{sec:conclusions} contains our conclusions.

%%%%%%%%%%%%%%%%%%%%%%%%%%%%%%%%%%%%%%%%%%%%%%%%%%%%%%%%%%%%%%%%%%%%%%%%%%%%%%%%
%%%%%%%%%%%%%%%%%%%%%%%%%%%%%%%%%%%%%%%%%%%%%%%%%%%%%%%%%%%%%%%%%%%%%%%%%%%%%%%%
\section{Validity of the assumptions}       \label{sec:assumptions}
%%%%%%%%%%%%%%%%%%%%%%%%%%%%%%%%%%%%%%%%%%%%%%%%%%%%%%%%%%%%%%%%%%%%%%%%%%%%%%%%

In the following, we will work with the double quark mass ratio $Q^2$
\beq
Q^2 = \frac{m_s - \hat{m}}{m_d - m_u} \frac{m_s + \hat{m}}{m_d + m_u}
%    = \frac{\sqrt{3}}{4 \epsilon} \frac{m_{K}^2}{m_{\pi}^2}
\eeq
instead of the mixing angle $\epsilon$, since the 
Kaplan-Manohar reparametrization invariance \cite{KM} of the chiral effective Lagrangian 
is respected by $Q^2$ up to chiral order $\mathcal{O}(p^4)$, whereas 
$\epsilon$ receives corrections already at $\mathcal{O}(p^2)$.
Hence, it is preferable to employ $Q$ in phenomenological analyses
in order to suppress the ambiguity stemming from this reparametrization invariance.

Following Dashen's theorem which asserts equal electromagnetic corrections
for pion and kaon masses at leading chiral order \cite{Dashen}, $Q^2$ can be expressed
in terms of physical meson masses
\beq
Q^2_{\textrm{Dashen}} = \frac{m_{K}^2}{m_{\pi}^2} \ \frac{m_{K}^2 -m_{\pi}^2}{
         m_{K^0}^{2} - m_{K^\pm}^{2} + m_{\pi^\pm}^{2} - m_{\pi^0}^{2}} = (24.1)^2 \ .
\eeq
However, there are various investigations on the size of violations to Dashen's theorem
with (partially contradictory) results for $Q$ in the range of about $21 \ldots 24$ \cite{CorrDash}.
The $3 \pi$ decays of $\eta$ and $\eta'$ provide thus a good opportunity to
pin down the value of the double quark mass ratio $Q^2$ \cite{Kambor, MS}.

We will first investigate the validity of assumption ``\ref{assuma}'', i.e.\ we assume that
the decay $\eta' \to \pi^0 \pi^+ \pi^-$ proceeds entirely 
via $\eta' \to \eta \pi^+ \pi^-$ followed by  $\pi^0$-$\eta$ mixing.
This implies for the neutral decay  
$A(\eta' \to 3 \pi^0) = 3 \epsilon \ A(\eta' \to \eta \pi^0 \pi^0)$.
Employing the amplitudes $A(\eta' \to \eta \,2 \pi)$ from the approach
advocated in \cite{BN1} --- which are in very good agreement with the data given
in \cite{pdg} --- one can thus predict the decay amplitudes for $A(\eta' \to 3 \pi)$
and calculate both the decay widths
$\Gamma(\eta' \to \pi^0 \pi^+ \pi^-)$, $\Gamma(\eta' \to 3 \pi^0)$ and the branching ratios
$r$ and $r_2 = \Gamma(\eta' \to 3 \pi^0)/\Gamma(\eta' \to \eta \pi^0 \pi^0)$.

In \cite{BN1} a least-squares fit to meson-meson scattering phase shifts
and $\eta, \eta'$ decays has been performed. One
observes four different classes of fits, i.e.\ clusters,
which describe these data equally well, but differ in their predictions for yet unmeasured
quantities such as the $\eta' \to \pi^0 \pi^+ \pi^-$ decay width (for which there exists
only a weak upper limit) and the Dalitz slope parameters of $\eta' \to 3 \pi$.
In fact, as we will see in the next section, inclusion of the recent VES data for
$\eta' \to \eta \pi^+ \pi^-$ \cite{VES2} reduces the number of fit clusters to one, 
but in the current section we take the data given in \cite{pdg}. We employ furthermore
the upper limit of 1.75\,\% for the branching fraction of $\eta' \to \pi^0 \pi^+ \pi^-$
as measured by the VES collaboration \cite{VES1} which is significantly lower than the 
previous upper limit of 5\,\% \cite{pdg}. This tighter bound translates to an upper limit 
of 3.8\,keV for the partial decay width and reduces the upper limit for $r$ from 
10\,\% (as quoted by the PDG) to 4.1\,\%.
The pertinent results for the four fit clusters of \cite{BN1} are well below these
new upper limits and can be utilized without modification.
As already reported in \cite{BN1}, the fit to the data does not allow for 
conclusions on the size of violations to Dashen's theorem
since $Q$ is treated as an input parameter and variations of $Q$ in the 
range of $20 \dots 24$ lead to equally good fits within our approach.
Hence, we will set $Q=24.1$ in our calculations---the value predicted by Dashen's theorem.
The results for the branching ratios obtained by 
employing assumption ``\ref{assuma}'' and $Q=24.1$ are shown in Table~\ref{tab:mix}.
The ratios are obtained by explicitly performing the integration of the
amplitudes over phase space.
Obviously, assumption ``\ref{assuma}'' is not justified---at least for the neutral decay
where, in particular, clusters 1 and 2 are in clear disagreement with experiment.

\begin{table}
\centering
\begin{tabular}{|l@{ }l||r@{$\,\pm\,$}l|r@{$\,\pm\,$}l|r@{$\,\pm\,$}l|r@{$\,\pm\,$}l||r@{$\,\pm\,$}l|}
\hline
& & \multicolumn{2}{c|}{Cluster 1} & \multicolumn{2}{c|}{Cluster 2}
& \multicolumn{2}{c|}{Cluster 3} & \multicolumn{2}{c||}{Cluster 4} & \multicolumn{2}{c|}{Exp.}\\
\hline
$\Gamma(\eta' \to \pi^0 \pi^+ \pi^-)$ & [eV]
&     69 &    12  &     73 &     9  &    141 &    44  &    141 &     26 & \multicolumn{2}{c|}{$< 3800$} \\
\hline
$r$ & [\%]
&   0.09 &  0.02  &   0.09 &  0.02  &   0.17 &  0.06  &   0.17 &   0.03 & \multicolumn{2}{c|}{$< 4.1$} \\
\hline
$\Gamma(\eta' \to 3 \pi^0)$ & [eV]
&    116 &    22  &    120 &    16  &    217 &    67  &    217 &     40 &    315 &     78 \\
\hline
$r_2$ & [\%]
&   0.26 &  0.05  &   0.26 &  0.04  &   0.47 &  0.15  &   0.47 &   0.08 &   0.74 &   0.12 \\
\hline
\end{tabular}
\caption{Decay widths and branching ratios in the chiral unitary approach \cite{BN1}
         employing assumption ``\ref{assuma}''.}
\label{tab:mix}
\end{table}

Next, we employ in addition assumption ``\ref{assumb}''. 
This is achieved by averaging the $\eta' \to \eta \,2 \pi$ amplitudes over phase space 
which are then employed for $\eta' \to 3 \pi$ by means of assumption ``\ref{assuma}''.
The results are displayed in Table~\ref{tab:mixconst}. One observes that for clusters~1 
and 2 the decay widths into $3 \pi$ and hence the ratios $r$, $r_2$ increase, while the 
changes for clusters~3 and 4 are rather moderate. However, recall that the Dalitz
plot parameters of the approach \cite{BN1} clearly indicate that
the assumption of a constant amplitude is not justified for $\eta' \to \pi^0 \pi^+ \pi^-$,
particularly for clusters~3 and 4.
The partial compensation of the effects of assumption ``\ref{assuma}'' in clusters~1, 2 
and the moderate changes in clusters~3, 4 are therefore purely accidental.

\begin{table}
\centering
\begin{tabular}{|l@{ }l||r@{$\,\pm\,$}l|r@{$\,\pm\,$}l|r@{$\,\pm\,$}l|r@{$\,\pm\,$}l||r@{$\,\pm\,$}l|}
\hline
& & \multicolumn{2}{c|}{Cluster 1} & \multicolumn{2}{c|}{Cluster 2}
& \multicolumn{2}{c|}{Cluster 3} & \multicolumn{2}{c||}{Cluster 4} & \multicolumn{2}{c|}{Exp.}\\
\hline
$\Gamma(\eta' \to \pi^0 \pi^+ \pi^-)$ & [eV]
&    155 &     7  &    155 &     7  &    153 &     7  &    154 &      5 & \multicolumn{2}{c|}{$< 3800$} \\
\hline
$r$ & [\%]
& \multicolumn{2}{c|}{0.19}  & \multicolumn{2}{c|}{0.19} & \multicolumn{2}{c|}{0.19} 
& \multicolumn{2}{c||}{0.19} & \multicolumn{2}{c|}{$< 4.1$} \\
\hline
$\Gamma(\eta' \to 3 \pi^0)$ & [eV]
&    238 &    11  &    239 &    10  &    237 &    11  &    239 &      6 &    315 &     78 \\
\hline
$r_2$ & [\%]
& \multicolumn{2}{c|}{0.52}  & \multicolumn{2}{c|}{0.52} & \multicolumn{2}{c|}{0.52}
& \multicolumn{2}{c||}{0.52} & 0.74 & 0.12 \\
\hline
\end{tabular}
\caption{Decay widths and branching ratios in the chiral unitary approach employing 
         assumptions ``\ref{assuma}'' and ``\ref{assumb}''.
         Since in this case the branching ratios only depend on phase space and $Q$,
         they do not have an error bar.}
\label{tab:mixconst}
\end{table}

We conclude that both assumptions ``\ref{assuma}'' and ``\ref{assumb}'' are not justified.
This is further substantiated by comparison of $r$ and $r_2$ in Tab.~\ref{tab:mixconst} 
with the respective values from the full chiral unitary approach shown in Tab.~\ref{tab:BN1}.
The values are in clear disagreement and, hence, both assumptions are not appropriate---at
least within the chiral unitary approach.

\begin{table}
\centering
\begin{tabular}{|l@{ }l||r@{$\,\pm\,$}l|r@{$\,\pm\,$}l|r@{$\,\pm\,$}l|r@{$\,\pm\,$}l||r@{$\,\pm\,$}l|}
\hline
& & \multicolumn{2}{c|}{Cluster 1} & \multicolumn{2}{c|}{Cluster 2}
& \multicolumn{2}{c|}{Cluster 3} & \multicolumn{2}{c||}{Cluster 4} & \multicolumn{2}{c|}{Exp.}\\
\hline
$\Gamma(\eta' \to \pi^0 \pi^+ \pi^-)$ & [eV]
&    470 &   200  &    520 &   200  &    740 &   420  &    620 &    180 & \multicolumn{2}{c|}{$< 3800$} \\
\hline
$r$ & [\%]
&   0.58 &  0.24  &   0.66 &  0.27  &   0.92 &  0.52  &   0.77 &   0.21 & \multicolumn{2}{c|}{$< 4.1$} \\
\hline
$\Gamma(\eta' \to 3 \pi^0)$ & [eV]
&    331 &    24  &    326 &    28  &    330 &    33  &    336 &     21 &    315 &     78 \\
\hline
$r_2$ & [\%]
&   0.73 &  0.06  &   0.72 &  0.06  &   0.71 &  0.07  &   0.73 &   0.05 &   0.74 &   0.12 \\
\hline
\end{tabular}
\caption{Decay widths and branching ratios in the chiral unitary approach \cite{BN1}.}
\label{tab:BN1}
\end{table}

Finally, we would like to investigate the differences which result if 
assumption ``\ref{assuma}'' is replaced by the decay mechanism where $\eta' \to 3 \pi$ 
occurs due to $\pi^0$-$\eta'$ mixing followed by a (virtual) transition $\pi^0 \to 3 \pi$.
Employing the relation $A(\eta' \to 3 \pi) = \epsilon' \ A(\pi^0 \to 3 \pi)$ with 
$\epsilon'$ being the $\pi^0$-$\eta'$ mixing angle \cite{BB1}
we find the values shown in Table~\ref{tab:etap_mix}.
Assuming the $\eta' \to 3 \pi$ decays to proceed via this mechanism introduces
a huge uncertainty and leads to different ratios $r$ and $r_2$.
This underlines the observation that the decays $\eta' \to 3 \pi$ cannot be 
expected to simply proceed either via $\pi^0$-$\eta$ or $\pi^0$-$\eta'$ mixing.
In particular, the isospin-breaking transition due to the quark mass difference $m_d - m_u$ 
cannot be completely assigned to $\pi^0$-$\eta$ mixing as done in assumption ``\ref{assuma}''. 
Despite its appealing simplicity, the crude estimate given in Eq.~(\ref{eq:gtw})
is certainly not suited to precisely determine the double quark mass ratio $Q^2$.
In fact, even at leading chiral order the $\eta' \to 3 \pi$ decay amplitude is
not entirely due to $\pi^0$-$\eta$ mixing. There is also a contribution from
an isospin-violating $\eta' 3 \pi$-vertex from the explicit chiral
symmetry breaking part of the Lagrangian at second chiral order, see e.g. Ref.~\cite{BB1}.

\begin{table}
\centering
\begin{tabular}{|l@{ }l||r@{$\,\pm\,$}l|r@{$\,\pm\,$}l|r@{$\,\pm\,$}l|r@{$\,\pm\,$}l||r@{$\,\pm\,$}l|}
\hline
& & \multicolumn{2}{c|}{Cluster 1} & \multicolumn{2}{c|}{Cluster 2}
& \multicolumn{2}{c|}{Cluster 3} & \multicolumn{2}{c||}{Cluster 4} & \multicolumn{2}{c|}{Exp.}\\
\hline
$\Gamma(\eta' \to \pi^0 \pi^+ \pi^-)$ & [eV]
&   2450 &  1930  &   1720 &  1160  &    260 &   260  &    290 &    290 & \multicolumn{2}{c|}{$< 3800$} \\
\hline
$r$ & [\%]
&   2.96 &  2.30  &   2.10 &  1.40  &   0.34 &  0.34  &   0.37 &   0.37 & \multicolumn{2}{c|}{$< 4.1$} \\
\hline
$\Gamma(\eta' \to 3 \pi^0)$ & [eV]
&   1080 &   840  &    800 &   550  &    120 &   120  &    120 &    120 &    315 &     78 \\
\hline
$r_2$ & [\%]
&   2.34 &  1.79  &   1.73 &  1.19  &   0.28 &  0.28  &   0.28 &   0.28 &   0.74 &   0.12 \\
\hline
\end{tabular}
\caption{Decay widths and branching ratios in the chiral unitary approach 
         if isospin-breaking takes place solely via $\pi^0$-$\eta'$ mixing.
         For the fits of Clusters~3 and 4 this mixing angle can actually
         become zero leading to vanishing decay widths $\Gamma(\eta' \to 3 \pi)$
         and branching ratios $r$, $r_2$.}
\label{tab:etap_mix}
\end{table}

On the other hand, employing the chiral unitary approach of \cite{BN1} does not 
lead to a conclusive extraction of $Q$ due to the present experimental situation. 
From a fit to the data in \cite{pdg}
supplemented by Dashen's theorem one obtains the decay width ratio $r=(0.35 \ldots 1.5)\%$
which is larger than the value of $0.18\%$ quoted in \cite{GTW}.
Note also that there is a tendency to even larger values of $r$ if $Q$ is lowered,
e.g., for $Q=22$ we obtain the range $r=(0.4 \ldots 2.8)\%$.

%%%%%%%%%%%%%%%%%%%%%%%%%%%%%%%%%%%%%%%%%%%%%%%%%%%%%%%%%%%%%%%%%%%%%%%%%%%%%%%%
%%%%%%%%%%%%%%%%%%%%%%%%%%%%%%%%%%%%%%%%%%%%%%%%%%%%%%%%%%%%%%%%%%%%%%%%%%%%%%%%
\section{Inclusion of the VES data for \boldmath$\eta' \to \eta \pi^+ \pi^-$}       \label{sec:VES}
%%%%%%%%%%%%%%%%%%%%%%%%%%%%%%%%%%%%%%%%%%%%%%%%%%%%%%%%%%%%%%%%%%%%%%%%%%%%%%%%

In this section we study the changes in our results that occur if the recent 
VES data on the spectral shape of $\eta' \to \eta \pi^+ \pi^-$ \cite{VES2} are taken 
into account. Note that the most recent analysis of the VES Collaboration \cite{VES2}
has not yet been included in Ref.~\cite{pdg}. 
The VES data have much higher statistics on the Dalitz slope 
parameters than previous experiments and by including them in the fit we obtain 
the results shown in Tab.~\ref{tab:VES}. 
Since the amplitudes for $\eta' \to \eta \pi^+ \pi^-$ and $\eta' \to \eta \pi^0 \pi^0$ 
are equal in the isospin limit and deviations are thus isospin-breaking and small 
in our approach, we only include the leading
Dalitz parameter $a$ of $\eta' \to \eta \pi^0 \pi^0$ \cite{Alde} and omit the higher ones
which are---assuming only small isospin-violating contributions---not quite compatible
with the new results of the VES experiment for $\eta' \to \eta \pi^+ \pi^-$. 
Our results are in good agreement with the Dalitz plot parameters extracted 
from the VES experiment.
In Tab.~\ref{tab:VES} only the best least-squares fit is shown which is sufficient
to discuss the qualitative changes of the results compared to those of 
Sec.~\ref{sec:assumptions}.
Note also that we have supplemented our fitting routine by a conjugate gradient
method \cite{GSL} and hence the numerical values have improved with respect to \cite{BN1}.

\begin{table}
\centering
\begin{tabular}{|l|c|c|c|}
\hline
\multicolumn{4}{|c|}{$\eta' \to \eta \pi^+ \pi^-$} \\
\hline
& $a$ & $b$ & $c$ \\
\hline
Theory &
$-0.109$ &
$-0.087$ &
$-0.036$ \\
\hline
Exp. &
$-0.127 \pm 0.024$  &
$-0.106 \pm 0.042$  &
$-0.082 \pm 0.025$  \\
\hhline{====}
\multicolumn{4}{|c|}{$\eta' \to \eta \pi^0 \pi^0$} \\
\hline
& $a$ & $b$ & $c$ \\
\hline
Theory &
$-0.123$ &
$-0.104$ &
$-0.041$ \\
\hline
Exp. &
$-0.116 \pm 0.026$  &
  &
  \\
\hline
\end{tabular}
\caption{Results for the Dalitz plot parameters of $\eta' \to \eta \pi \pi$ if the VES data 
         are included in the fit.}
\label{tab:VES}
\end{table}

Our results for the $\eta' \to 3 \pi$ decay widths and width ratios
are displayed in Tab.~\ref{tab:mixVES}. It is important to emphasize that
the inclusion of the VES data reduces the number of fit clusters to one and
we observe indeed one global minimum.
There is, however, a strong tendency of the fits towards the upper limit
$\Gamma(\eta' \to \pi^0 \pi^+ \pi^-) < 3.8\,$keV and, in fact, slightly improved fits
with a smaller $\chi^2$ value can be obtained if this upper limit is omitted.
(In this case, the best overall fit leads to the width
$\Gamma(\eta' \to \pi^0 \pi^+ \pi^-) = 5.73\,$keV.)
The result for the width ratio, $r=3.9\%$, has thus increased if the VES data are taken into account.
\begin{table}
\centering
\begin{tabular}{|l@{ }l||r||r@{$\,\pm\,$}l|}
\hline
& & \multicolumn{1}{c||}{Theory}
& \multicolumn{2}{c|}{Exp.}\\
\hline
$\Gamma(\eta' \to \pi^0 \pi^+ \pi^-)$ & [eV]
&     3120  & \multicolumn{2}{c|}{$< 3800$} \\
\hline
$r$ & [\%]
&     3.9   & \multicolumn{2}{c|}{$< 4.1$} \\
\hline
$\Gamma(\eta' \to 3 \pi^0)$ & [eV]
&       330  & 315 & 78 \\
\hline
$r_2$ & [\%]
&     0.73  & 0.74 & 0.12 \\
\hline
\end{tabular}
\caption{Decay widths and branching ratios if the VES data are taken into account.}
\label{tab:mixVES}
\end{table}
Furthermore, the amplitude $A(\eta' \to \pi^0 \pi^+ \pi^-)$
fluctuates strongly over phase space with slope parameters which can be more
than one order of magnitude larger in size than those obtained in \cite{BN1}.
The Dalitz plot distribution $|A(\eta' \to \pi^0 \pi^+ \pi^-)|^2$
can therefore not be properly described by a low-order polynomial in the usual expansion variables
$x$ and $y$, see \cite{BN1} for definitions, and it can definitely not be assumed
to be constant over phase space.

\begin{figure}
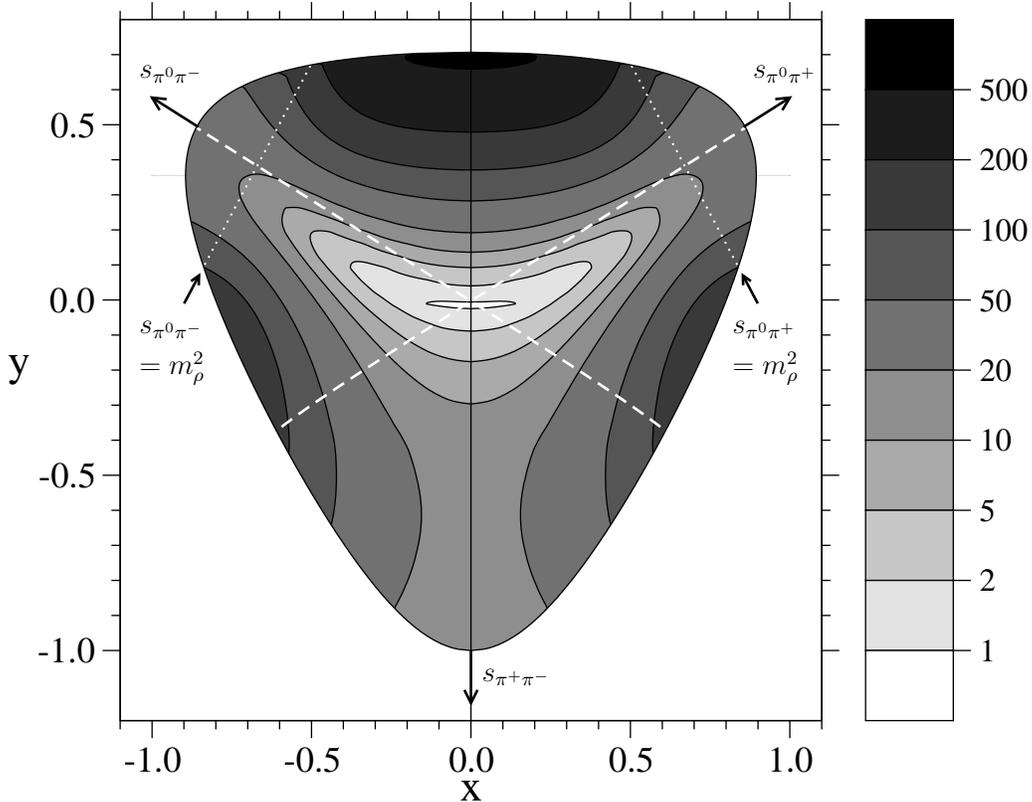

\centering
\begin{overpic}[width=0.8\textwidth]{DalVES.eps}
\put(46.5,12){\scalebox{0.9}{$s_{\pi^+ \pi^-}$}}
\put(73,71){\scalebox{0.9}{$s_{\pi^0 \pi^+}$}}
\put(13,71){\scalebox{0.9}{$s_{\pi^0 \pi^-}$}}
\put(71,46){\scalebox{0.9}{$s_{\pi^0 \pi^+}$}}
\put(71,42){\scalebox{0.9}{$= m^{2}_{\rho}$}}
\put(13,46){\scalebox{0.9}{$s_{\pi^0 \pi^-}$}}
\put(13,42){\scalebox{0.9}{$= m^{2}_{\rho}$}}
\end{overpic}
\caption{Dalitz plot distribution $|A(\eta' \to \pi^0 \pi^+ \pi^-)|^2$ of the best overall fit including
         the VES data for $\eta' \to \eta \pi^+ \pi^-$ \cite{VES2}. The distribution is 
         normalized to unity at $x = y = 0$ (see \cite{BN1} for definitions of $x$, $y$).
         The $p$-wave contributions to $\pi^0 \pi^+$ ($\pi^0 \pi^-$) 
         rescattering vanish on the rising (falling) dashed line
         and the invariant energies associated with the $\rho^{\pm}(770)$ are indicated by
         the dotted lines.}
\label{fig:Dal}
\end{figure}

The reason for both the large decay width $\Gamma(\eta' \to \pi^0 \pi^+ \pi^-)$ 
and the strong fluctuations over phase space are mainly due to a large contribution
from isospin $I=1$ $p$-waves in the final-state interactions of the decay.
While for $I=1$ $p$-waves the uncharged two-particle channels are $C$-even and, 
due to $C$-invariance, do not couple to $C$-odd channels related to the $\rho^0(770)$ as 
already pointed out in \cite{BN1}, the coupling of charged channels to 
the $\rho^{\pm}(770)$ is 
not forbidden.  In fact, an important feature of the fits including the
VES data compared to those without these is the large enhancement of the 
$\eta' \pi^{\pm} \to \pi^0 \pi^{\pm}$ coupling which also determines the importance of the 
$\rho^{\pm}(770)$ in this decay. The pertinent Dalitz plot is shown in Fig.~\ref{fig:Dal}
and exhibits signatures of the $\rho^{\pm}(770)$. Note, however, that these resonances
do not appear as bands of increased amplitude at fixed two-particle energies
(dotted lines in Fig.~\ref{fig:Dal}), since the $p$-wave contributions have a kinematical 
zero in the middle of these bands as indicated in Fig.~\ref{fig:Dal} (dashed lines). 
Thus the amplitude only peaks at the edge of the Dalitz plot. 
Moreover, due to the symmetry of the amplitude under $\pi^+ \leftrightarrow \pi^-$ exchange
($C$-invariance) the $\rho^+$, $\rho^-$ peaks interfere constructively on the symmetry
axis producing a pronounced peak structure at the top of the Dalitz plot, where the 
invariant mass of the $\pi^+ \pi^-$ system is minimal.
These features of the Dalitz plot of a pseudoscalar meson decaying into three pions have 
been pointed out long ago in \cite{Zem}.

%%%%%%%%%%%%%%%%%%%%%%%%%%%%%%%%%%%%%%%%%%%%%%%%%%%%%%%%%%%%%%%%%%%%%%%%%%%%%%%%
%%%%%%%%%%%%%%%%%%%%%%%%%%%%%%%%%%%%%%%%%%%%%%%%%%%%%%%%%%%%%%%%%%%%%%%%%%%%%%%%
\section{Conclusions}       \label{sec:conclusions}
%%%%%%%%%%%%%%%%%%%%%%%%%%%%%%%%%%%%%%%%%%%%%%%%%%%%%%%%%%%%%%%%%%%%%%%%%%%%%%%%

In this work, we have critically investigated the claim of Ref.~\cite{GTW}
that the light quark mass ratio $(m_d - m_u)/(m_s - \hat{m})$ can be
extracted from the decay width
ratio $r = \Gamma(\eta' \to \pi^0 \pi^+ \pi^-)/\Gamma(\eta' \to \eta \pi^+ \pi^-)$.
In order to study this issue we have employed a U(3) chiral unitary framework
developed in \cite{BB1, BN1} which is in very good agreement with the $\eta$, $\eta'$ data
on widths and spectral shapes.
Our results clearly indicate that the two underlying assumptions of \cite{GTW} in order 
to arrive at a relation between $r$ and $(m_d - m_u)/(m_s - \hat{m})$, i.e.,
that \ref{assuma}) the decay $\eta' \to \pi^0 \pi^+ \pi^-$ proceeds entirely 
via the decay $\eta' \to \eta \pi^+ \pi^-$ followed by  $\pi^0$-$\eta$ mixing
and that \ref{assumb})  the decay amplitudes are constant over phase space, are not justified at all.
The results from the full chiral unitary approach are in plain disagreement with these
two assumptions. Moreover, the present experimental situation which
is used as input to fit the parameters of the chiral unitary approach does
not allow for a precise determination of the double quark mass ratio $Q^2$
from $r$.

Inclusion of the recent VES data on the $\eta' \to \eta \pi^+ \pi^-$ spectral shape 
reduces the uncertainty of the fit results to some extent. In this case, 
the overall fit to $\eta$, $\eta'$ data yields for $\eta' \to \pi^0 \pi^+ \pi^-$ 
a large contribution from the isospin $I=1$ $p$-wave in the final-state 
interactions which can be attributed to a large coupling to the $\rho^{\pm}(770)$ 
resonances while contributions related to the $\rho^0(770)$ are forbidden by 
$C$-invariance.
More precise data on $\eta$ and $\eta'$ decays are needed in order to eventually
clarify this issue. An improvement of the experimental situation
is foreseen in the near future due to the upcoming data from WASA at COSY \cite{WASA},
MAMI-C \cite{Nef} and KLOE at DA$\Phi$NE \cite{KLOE}.

%%%%%%%%%%%%%%%%%%%%%%%%%%%%%%%%%%%%%%%%%%%%%%%%%%%%%%%%%%%%%%%%%%%%%%%%%%%%%%%%
\section*{Acknowledgments}
We thank Heiri Leutwyler for reading the manuscript and useful comments.
This research is part of the EU Integrated Infrastructure Initiative Hadron Physics Project
under contract number RII3-CT-2004-506078. Work supported in part by DFG (SFB/TR 16,
``Subnuclear Structure of Matter'', and BO 1481/6-1).

%%%%%%%%%%%%%%%%%%%%%%%%%%%%%%%%%%%%%%%%%%%%%%%%%%%%%%%%%%%%%%%%%%%%%%%%%%%%%%%%

\end{document}